\documentstyle[12pt]{article}
\begin{document}
\begin{center}

{\Large Vanishing of the Bare Coupling in Four Dimensions}
\vskip .8cm

V. {Elias $^{a,b,}$}\footnote{Electronic address: {\tt velias@uwo.ca}} 
and D.G.C. {McKeon $^{b,c,}$}\footnote{Electronic address: {\tt dgmckeo2@uwo.ca}}
 
\baselineskip=12pt

\vskip .5cm
{\small \it
$^{a}$ 
Perimeter Institute for Theoretical Physics, 35 King Street North,\\
Waterloo, Ontario  N2J 2W9 CANADA}\\
{\small \it
$^{b}$ 
Department of Applied Mathematics, The University of Western Ontario,\\
London, Ontario  N6A 5B7 CANADA}\\
{\small \it
$^{c}$ 
Department of Mathematical Physics, National University of Ireland,\\
Galway, IRELAND}
\end{center}

\begin{abstract}We examine two restructurings of the series relationship between the bare and 
renormalized coupling constant in dimensional regularization.  In one of these restructurings, 
we are able to demonstrate via all-orders summation of leading and successive $\epsilon = 0$ 
(dimensionality = 4) poles that the bare coupling vanishes in the dimension-4 limit.
\end{abstract}

\newpage

\setcounter{footnote}{0}

\baselineskip=24pt

In the context of dimensional regularization \cite{1} with minimal subtraction, 
the bare $(g_B)$ and renormalized $(g)$ coupling constants are related by a series 
of the form \cite{2,3}

\begin{eqnarray}
g_B & = & \mu^\epsilon g\left[1 + a_{1,1} g^2/\epsilon + a_{2,1}
g^4/\epsilon + a_{2,2} g^4/\epsilon^2 + a_{3,1} g^6/\epsilon + a_{3,2}
g^6/\epsilon^2 + a_{3,3} g^6/\epsilon^3 + \ldots\right]\nonumber\\
& = & \mu^\epsilon \sum_{\ell = 0}^\infty \sum_{k=\ell}^\infty
a_{k,\ell} g^{2k+1} \epsilon^{-\ell},
\end{eqnarray}
where $a_{k,0} \equiv \delta_{k,0}$ and $\epsilon \equiv 2 - n/2$ in $n$
dimensions. This double summation is meaningful 
provided $g$ is sufficiently small and provided $\epsilon^{-1}$ is
finite. On the basis of eq. (1), however it is generally held that 
the bare coupling becomes infinite in the 4-dimensional limit
(i.e. $g_B 
\rightarrow \infty$ as $\epsilon \rightarrow 0$).  We argue in this
note that renormalization group $(RG)$ methods may be used to show
$\displaystyle{\lim_{n \rightarrow 4} g_B = 0}$, 
a result consistent with asymptotic-freedom
expectations in which the bare coupling is the renormalized coupling in
the infinite cut-off limit.

We begin first by noting that the series (1) may be reorganized as
follows:
\begin{equation}
g_B = \mu^\epsilon g\sum_{n=0}^\infty g^{2n}
S_n\left(g^2/\epsilon\right)
\end{equation}
where the functions $S_n (u)$ have power series expansions
\begin{equation}
S_n(u) \equiv \sum_{m=n}^\infty a_{m,m-n}u^{m-n}.
\end{equation}
In particular, the summation over leading order poles is just
\begin{equation}
S_0\left(g^2/\epsilon\right) = a_{0,0} + a_{1,1} g^2/ \epsilon +
a_{2,2}\left(g^2/\epsilon\right)^2 + \ldots,
\end{equation}
and that over next-to-leading poles is
\begin{equation}
gS_1\left(g^2/\epsilon\right) = a_{1,0} \; g + a_{2,1} \; g^3/ \epsilon +
a_{3,2} \; g^5/\epsilon^2 + \ldots.
\end{equation}

Explicit summation of the series $S_n(u)$ is possible given full knowledge of 
the $\beta$-function
characterising the scale dependence of the coupling constant $g$.
Since $g_B$ is independent of the mass-scale $\mu$ introduced for
dimensional consistency [i.e. the action $\int d^n x \times {\cal{L}}$ must be
dimensionless], we find that in $n$-dimensions \cite{3}
\begin{eqnarray}
\mu \frac{dg_B}{d\mu} = 0 = \left( \mu\frac{\partial}{\partial \mu} +
\tilde{\beta} (g) \frac{\partial}{\partial g}\right) g_B\nonumber\\
= \epsilon g_B + \left(-\epsilon + \sum_{n=1}^\infty b_{2n+1} \; 
g^{2n}\right) g \frac{\partial g_B}{\partial g}
\end{eqnarray}
where $\tilde{\beta}(g) (= -\epsilon g + \beta(g))$ turns into the usual $\beta$
function series in the limit $\epsilon \rightarrow 0$:
\begin{equation}
\mu \frac{dg}{d\mu} \equiv \tilde{\beta} (g) 
\longrightarrow \sum_{n=1}^\infty b_{2n+1} \; g^{2n+1}.
\end{equation}
Upon substituting eq. (1) into eq. (6), we find that
\begin{eqnarray}
0 = \sum_{n=1}^\infty \left(\frac{1}{\epsilon}\right)^{n-1} \sum_{m=n}^\infty (-2m)a_{m,n}
g^{2m+1}\nonumber\\
+ \sum_{\ell = 0}^\infty \left(\frac{1}{\epsilon}\right)^\ell \sum_{p = \ell}^\infty (2p+1)
a_{p,\ell} \sum_{q=1}^\infty b_{2q +1} g^{2(p+q)+1}.
\end{eqnarray}
Such a relationship between $\beta$-function coefficients and coefficients of poles in
eq. (1) has been noted by 't Hooft \cite{2} and by Collins and Macfarlane \cite{3}.  
In eq. (8), the aggregate coefficient of $g^{2\ell + 1}(1/\epsilon)^{\ell - 1}$ is just
\begin{equation}
-2\ell a_{\ell , \ell} + b_3(2\ell - 1) a_{\ell - 1, \ell - 1} = 0,
\; \; \; \ell \geq 1.
\end{equation}
Ordinarily, one would use this equation to obtain $b_3$ from the
calculated value of $a_{1,1}$ ($a_{0,0} = 1$, and $b_3 = 2a_{1,1}$).
However, we see that this recursion relation also determines all coefficients
$a_{m,m}$ within the summation (4) over leading-order poles.
Similarly, the aggregate coefficient of $g^{2\ell +1}(1/\epsilon)^{\ell
- 2}$ within eq. (8),
\begin{equation}
-2\ell a_{\ell , \ell - 1} + b_3 (2\ell -1) a_{\ell - 1, \ell - 2} + b_5(2\ell - 3)
a_{\ell - 2, \ell - 2} = 0, \; \; \ell \geq 2,
\end{equation}
not only implies $b_5 = +4a_{2,1}$,  $(a_{1,0} = 0)$, but also determines all coefficients
$a_{m, m-1}$ within the summation (5) of next-to-leading-order poles. Indeed, the aggregate
coefficient of $g^{2\ell + 1} (1/\epsilon)^{\ell - k}$ within eq. (8),
\begin{equation}
-2\ell a_{\ell, \ell - k + 1} + \sum_{q=1}^k b_{2q+1} (2\ell - 2q + 1)
a_{\ell - q, \ell - k} = 0, \; \; \ell \geq k \geq 1
\end{equation}
serves as a recursion relation for the evaluation of $S_{k-1} (u)$, as defined
by the summation (3).

To evaluate explicitly the summations $S_n$ within eq. (2), we begin by
multiplying the recursion relation (9) by $u^{\ell - 1}$ and summing
from $\ell = 1$ to infinity:
\begin{eqnarray}
0 = -2 \sum_{\ell = 1}^\infty \ell a_{\ell , \ell} u^{\ell - 1} + b_3
\sum_{\ell = 1}^\infty (2\ell - 1)a_{\ell - 1, \ell - 1}u^{\ell -
1}\nonumber\\
= -2 \frac{dS_0(u)}{du} + 2b_3 u\frac{dS_0(u)}{du} + b_3
S_0(u).
\end{eqnarray}
The final line of eq. (12) is obtained using the definition (4) for
$S_0(u)$.  Moreover, we see from eq. (4) that $S_0(0) = a_{0,0} = 1$, in
which case the solution to the separable first-order differential
equation (12) is just
\begin{equation}
S_0(u) = (1 - b_3 u)^{-1/2}.
\end{equation}
Similarly, we can obtain a differential equation for $S_1(u)$ by
multiplying the recursion relation (10) by $u^{\ell - 2}$ and then
summing from $\ell = 2$ to $\infty$.  One easily finds from the
definitions (3) of $S_0$ and $S_1$ that
\begin{eqnarray}
2\left(1 - b_3 u\right) \frac{dS_1}{du} + \left(\frac{2}{u} -
3b_3\right)S_1\nonumber\\
= b_5 \left[2u \frac{dS_0}{du} + S_0\right]
\end{eqnarray}
where $S_0(u)$ is given by (13), and where $S_1(0) = a_{1,0} = 0$.
Upon making a change in variable to
\begin{equation}
w = 1 - b_3u,
\end{equation}
one finds after a little algebra that
\begin{equation}
\frac{dS_1}{dw} + \frac{(1-3w)}{2w(1-w)} S_1 = -\frac{b_5}{2b_3} w^{-
5/2},
\end{equation}
with initial condition $S_1|_{w=1} = S_1|_{u=0} = 0$. The solution of
this differential equation is
\begin{equation}
S_1[w(u)] = -\frac{b_5}{2b_3 w^{1/2}(w-1)}\left(\log w + \frac{1}{w} -
1\right).
\end{equation}
It is worthwhile to note that
\begin{equation}
S_0 = w^{-1/2}
\end{equation}
and that as $w \rightarrow \infty$,
\begin{equation}
S_1 \longrightarrow -\frac{b_5}{2b_3}
w^{-3/2} \log (w).
\end{equation}
For an asymptotically-free theory $(b_3 < 0)$, the $w \rightarrow \infty$ limit corresponds to the limit $\epsilon
\rightarrow 0^+$ when $u = g^2/\epsilon$.  Thus the restriction $w = 1 -
b_3 g^2 / \epsilon > 0$ on the domain of eq. (17) as a real function
necessarily implies for asymptotically free theories that the
dimensionality $n = 4 - 2\epsilon$ approaches four from below.

Consider now the general relation (11), where the index $k$ is taken to
be greater than or equal to $2$.  If we multiply Eq. (11) by $u^{\ell
- k}$ and then sum from $\ell = k$ to $\infty$, we obtain the following
differential equation via the definition (3):
\begin{eqnarray}
0 & = & -2 \sum_{\ell = k}^\infty \left[(\ell -k + 1) + (k - 1) \right]
a_{\ell , \ell - k + 1} u^{\ell - k}\nonumber\\
& + & \sum_{q=1}^k b_{2q+1} \sum_{\ell = k}^\infty \left[2(\ell - k) +2(k-q) +
1\right] a_{\ell - q, \ell - k}u^{\ell -k}\nonumber\\
& = & -2\frac{dS_{k-1}}{du} - \frac{2(k-1)}{u} S_{k-1}
+ \sum_{q=1}^k b_{2q+1} \left[ 2u \frac{dS_{k-q}}{du} + (2(k-q)+1)
S_{k-q}\right].
\end{eqnarray}
By letting $k - 1 \rightarrow k$ and then making use of the 
change-of-variable (15), we obtain the following differential equation 
for $S_k[w]$ with $k \geq 2$:
\begin{eqnarray}
& &\frac{dS_k}{dw} + \left[\frac{(2k+1)w-1}{2w(w-1)} \right] S_k \nonumber\\
& = & -\frac{b_{2k+3}}{2b_3} w^{-5/2}
- \frac{1}{2b_3} \sum_{n=1}^{k-1} \frac{b_{2n+3}}{w}\left[2(w-
1)\frac{dS_{k-n}}{dw} + (2k - 2n + 1)S_{k-n}\right].
\end{eqnarray}
The first term on the right hand side of eq. (21) is obtained making
explicit use of the expression (18) for $S_0$. Note that if $k \geq 1$, $S_k(u=0) =
S_k[w=1] = 0$. Since $S_0$ and $S_1$ are already
known, one may solve the $k = 2$ case of eq. (21) for $S_2$, then use this
solution within the $k = 3$ version of eq. (21) to solve for $S_3$, etc.,
so as to obtain {\it{all}} $S_k$ explicitly.\footnote{Such an iteration
of solutions is possible only if all coefficients $b_{2n+3} =
2(n+1)a_{n+1,1}$ are already known.}  Given knowledge of $\left\lbrace
S_{k-1}, S_{k-2}, \ldots , S_0\right\rbrace$, one finds the solution to
the differential equation (21) for $S_k$ to be
\begin{equation}
S_k[w] = \frac{
-\frac{1}{2b_3} \int_1^w dr\,r^{1/2} (r-1)^k\left[\sum_{n=1}^k
\frac{b_{2n+3}}{r}\left[2(r-1)\frac{d}{dr} + 2(k-n)+1\right]S_{k-
n}[r]\right]}{w^{1/2}(w-1)^k}.
\end{equation}
We note from eqs. (17) and (18) that
\begin{equation}
\frac{1}{r} \left[ 2(r-1)\frac{d}{dr} + 1\right]S_0[r] = r^{-
5/2}
\end{equation}
\begin{equation}
\frac{1}{r} \left[ 2(r-1)\frac{d}{dr} + 3\right]S_1[r] =-\frac{b_5}{b_3} 
\left(r^{-5/2} + {\cal{O}}\left(r^{-7/2}\right)\right).
\end{equation}
Consequently, one easily finds from eq. (22) that in the large $w$ limit
\begin{equation}
S_2 \sim w^{-3/2}.
\end{equation}
Upon substituting this behaviour into Eq. (22) for the $k = 3$ case, we
then find that as $w \rightarrow \infty$, $S_3 \sim w^{-3/2}$, in which
case successive iterations of Eq. (22) in the large $w$ limit
necessarily reproduce this asymptotic behaviour, regardless of the index $k$:
\begin{equation}
S_k[w] \sim w^{-3/2}, \;\; k \geq 2.
\end{equation}
We then see from eqs. (18), (19) and (26) that for {\it all} $k$,
\begin{equation}
\lim_{\epsilon \rightarrow 0} S_k\left(\frac{g^2}{\epsilon}\right) =
\lim_{w\rightarrow \infty} S_k[w] = 0,
\end{equation}
which implies via eq. (2) that
\begin{equation}
\lim_{\epsilon \rightarrow 0} g_B = 0.
\end{equation}

Recall in a cut-off regularization
scheme that one defines the bare coupling $g_B$ to be the infinite cut-off limit of
the renormalized coupling $g_R$:
\begin{equation}
g_B = \lim_{\Lambda \rightarrow \infty} g_R (\Lambda) = 0.
\end{equation}
The result (28) confirms for asymptotically free theories
that the infinite cut-off limit in four
dimensions and the $n = 4$ limit within dimensional regularization are
consistent.\footnote{We are grateful to V. A. Miransky for pointing out that
the result (29) is formally correct for non-asymptotically-free
theories ($b_3 > 0$) as well.  To one-loop order $g_R^2(\Lambda) 
= g_R^2 (\mu_0) / \left[ 1 - 2 b_3 g_R^2 (\mu_0) \log (\Lambda/\mu_0) \right].$  
If $b_3 > 0$, $g_R^2 (\Lambda)$ still goes to zero as $\Lambda \rightarrow \infty$.  However,
such large-$\Lambda$ behaviour is on the unphysical $\left( g_R^2 (\Lambda) < 0 \right)$ side 
of this expression's Landau pole. Consequently, the infinite cut off limit is
inappropriate for non-asymptotically free theories, as discussed in ref. \cite{4}. An
alternate discussion of how the bare coupling behaves in $\phi_4^4$ theory appears in ref. \cite{9}.\\}
Moreover, we note that the results (18), (19) and (26), which lead to Eq. (28), are not contingent
upon the sign of $b_3$;  the bare coupling constant is seen to vanish in the four-dimensional limit
even if the theory is {\it not} asymptotically free $\left( \epsilon \rightarrow 0^- \right)$.
Thus, the infinities which occur order-by-order in an expansion such as (1) disappear upon all-orders
summation of the $S_n (g^2/\epsilon)$ sub-series (3) within the reorganized expansion (2).\footnote{The use 
of RG-invariance to obtain such all-orders summations has been applied to other perturbative processes \cite{5}.}

Finally, we note that the series (1) may be reorganized into a power series in $\epsilon^{-1}$,
\begin{equation}
g_B = \mu^\epsilon \sum_{k=0}^\infty B_k (g) \epsilon^{-k},
\end{equation}
where
\begin{equation}
B_0 (g) = g,
\end{equation}
\begin{equation}
B_k (g) = \sum_{\ell = k}^\infty a_{\ell, k} g^{2\ell + 1}, \; \; \; k \geq 1 .
\end{equation}
The RG-invariance of $g_B$ to changes in $\mu$ may be utilised to determine the coefficients $B_k (g)$ explicitly.
The requirement that
\begin{eqnarray}
0 = \mu \frac{dg_B}{d\mu} & = & \mu^\epsilon \left[ \epsilon B_0 + \sum_{k=1}^\infty B_k (g) \epsilon^{1-k} \right.\nonumber\\
& + & \left. \tilde{\beta} (g) \sum_{k=0}^\infty \frac{d B_k}{dg} \epsilon^{-k} \right],
\end{eqnarray}
with $B_0 = g$ and $\tilde{\beta}(g) = -\epsilon g + \beta (g)$, as before, leads to a 
differential recursion relation \cite{6}
\begin{equation}
\left( \frac{d}{dg} - \frac{1}{g} \right) B_{k+1} (g) = \frac{1}{g} \beta(g) \frac{d B_k}{dg}.
\end{equation}
Noting from eqs. (31) and (32) that $B_k (0) = 0$, we solve eq. (34) to obtain
\begin{equation}
B_{k+1} (g) = g \int_0^g \frac{\beta(s) B_k^\prime (s)}{s^2} ds,
\end{equation}
where $\beta(s) = \sum_{n=1}^\infty b_{2n+1} s^{2n+1}$ as before.  We then find from
eqs. (31) and (35) that
\begin{equation}
B_1 (g) = \sum_{n=1}^\infty b_{2n+1} g^{2n+1} / 2n ,
\end{equation}
\begin{eqnarray}
B_2 (g) & = & \sum_{n=1}^\infty \sum_{k=1}^\infty \frac{b_{2n+1} b_{2k+1} (2k+1)}{4k(n+k)} g^{2(n+k+1)}\nonumber\\
& = & \frac{3b_3^2 g^5}{8} + \frac{11b_3 b_5 g^7}{24} + \left( \frac{8b_3 b_7}{3} + \frac{5b_5^2}{4} \right) \frac{g^9}{8} + ... \; ,
\end{eqnarray}
\begin{equation}
B_3 (g) = \frac{5 b_3^3 g^7}{16} + \frac{7b_3^2 b_5 g^9}{12} + ... \; .
\end{equation}
Of course, the process of iterating eq. (35) can be repeated to obtain $B_k (g)$ for arbitrarily large $k$, assuming
as before that all $\beta$-function coefficients $b_{2n+1}$ are known.  
For example, the $\beta$-function for $N = 1$ supersymmetric Yang-Mills theory can be extracted to all orders
via imposition of the Adler-Bardeen theorem on the anomaly supermultiplet \cite{7}, or via instanton
calculus methods \cite{8}; however, eq. (31) may no longer be valid since minimal subtraction
is not explicit in either of these approaches. Alternatively, one can show via eq. (36)
that terms of order $\epsilon^{-1}$ in eq. (1) are sufficient in themselves to determine $\beta(g)$ [see footnote 1],
or even that $\beta$-function coefficients $b_{2\ell+1}$ can be extracted via Eq. (37) provided
$g_B$ is known to order $g^{2\ell + 3}/\epsilon^2$.

\section*{Acknowledgements}

We are grateful for discussions with M. Davison, V. A. Miransky and P. J. Sullivan, and for financial support from the 
Natural Sciences and Engineering Research Council of Canada.

\end{document}